# Fake News Detection using Semi-Supervised Graph Convolutional Network

Priyanka Meel, Dinesh Kumar Vishwakarma
Biometric Research Laboratory, Department of Information Technology
Delhi Technological University, New Delhi, India-110042

**Abstract:** Social media becomes the central way for people to obtain and utilise news, due to its rapidness and inexpensive value of data distribution. Though, such features of social media platforms also present it a root cause of fake news distribution, causing adverse consequences on both people and culture. Hence, detecting fake news has become a significant research interest for bringing feasible real time solutions to the problem. Most current techniques of fake news disclosure are supervised, that need large cost in terms of time and effort to make a certainly interpreted dataset. The proposed framework concentrates on the text-based detection of fake news items while considering that only limited number of labels are available. Graphs are functioned extensively under several purposes of real-world problems on the strength of their property to structure things easily. Deep neural networks are used to generate great results within tasks that utilizes graph classification. The Graph Convolution Network works as a deep learning paradigm which works on graphs. Our proposed framework deals with limited amount of labelled data; we go for a semi-supervised learning method. We come up with a semi-supervised fake news detection technique based on GCN (Graph Convolutional Networks). The recommended architecture comprises of three basic components: collecting word embeddings from the news articles in datasets utilising GloVe, building similarity graph using Word Mover's Distance (WMD) and finally applying Graph Convolution Network (GCN) for binary classification of news articles in semi-supervised paradigm. The implemented technique is validated on three different datasets by varying the volume of labelled data achieving 95.27 % highest accuracy on Real or Fake dataset. Comparison with other contemporary techniques also reinforced the supremacy of the proposed framework.

## 1. Introduction

The constant extension of social media presented multiple parallel channels for users with added versatile facilities to obtain news than a few decades back. In accordance with the Pew

Research Center [1] , nearly 2/3 of U.S. grown-ups made headlines of social media. Since people proceed to profit of the service and simple convenience from social media, they additionally reveal themselves to some captivating and incorrect news published on social media platforms particularly fake knowledge. These fake news stories are purposely composed to carry incorrect information as a kind of idea so as business or governmental manipulation. The widespread of fake headlines could induce severe consequences on people and culture. Thus, identifying and alleviating fake news becomes a critical obstacle in current social media themes. The term "fake news" can be described in varied configurations that slightly differs from each other but most of the times can be used interchangeably as: misinformation, disinformation, hoax, propaganda, clickbait, misconception, fabrication, deceit etc. [2] [3] [4] [5] as pictured in word cloud representation in Figure 1.

The ubiquitous nature of web platforms encouraged their overwhelming use that fosters information sharing in an unrestricted way. This uncontrolled nature usually leads to generation and dissemination of biased contents. These untruthful events cause damage to social stability, economy and politics. One such rumour circulated on social media on 30 may 2018 regarding the spread of Nipah Virus through broiler chicken; causing huge financial loss to the dealers. On 4 January 2018 a fake news regarding collapse of metro pillar in Bangalore created unrest and panic in the public. A tidal wave of false claims and remedies has been burst out following COVID19 pademic such as curing the infection with garlic water, inhaling hot hair with hairdryer, use of certain vitamins or snake oils as remedy, use of vodka or vinegar as most effective hand sanitizer, unleashing lions on Russian streets to keep people indoor etc. These are few of the numerous instances of fake news circulating like wild fire on social media every day.

Paramount research in literature is engrossed around detecting fraudulent contents using rule-based methods or traditional machine learning classifiers. These traditional techniques fail to provide exemplary performance in the presence of large volumes of data, suffers with annotation inconsistencies, necessitate huge efforts for feature engineering and cannot provide near real time solutions. The reasons of inaccurate fact circulation spans across a broad spectrum incorporating genuine reporting faults, false interpretations, incorrect use of idioms or phrases, monetary benefits, political propaganda, marketing gimmick etc. but in all its camouflaged formats it poses a serious threat in front of the mankind. Therefore, in this work we tried to come up with a solution to this problem taking into consideration three key technologies involving GloVe word embedding, Word Mover's Distance (WMD) and Graph Convolutional Networks (GCN).

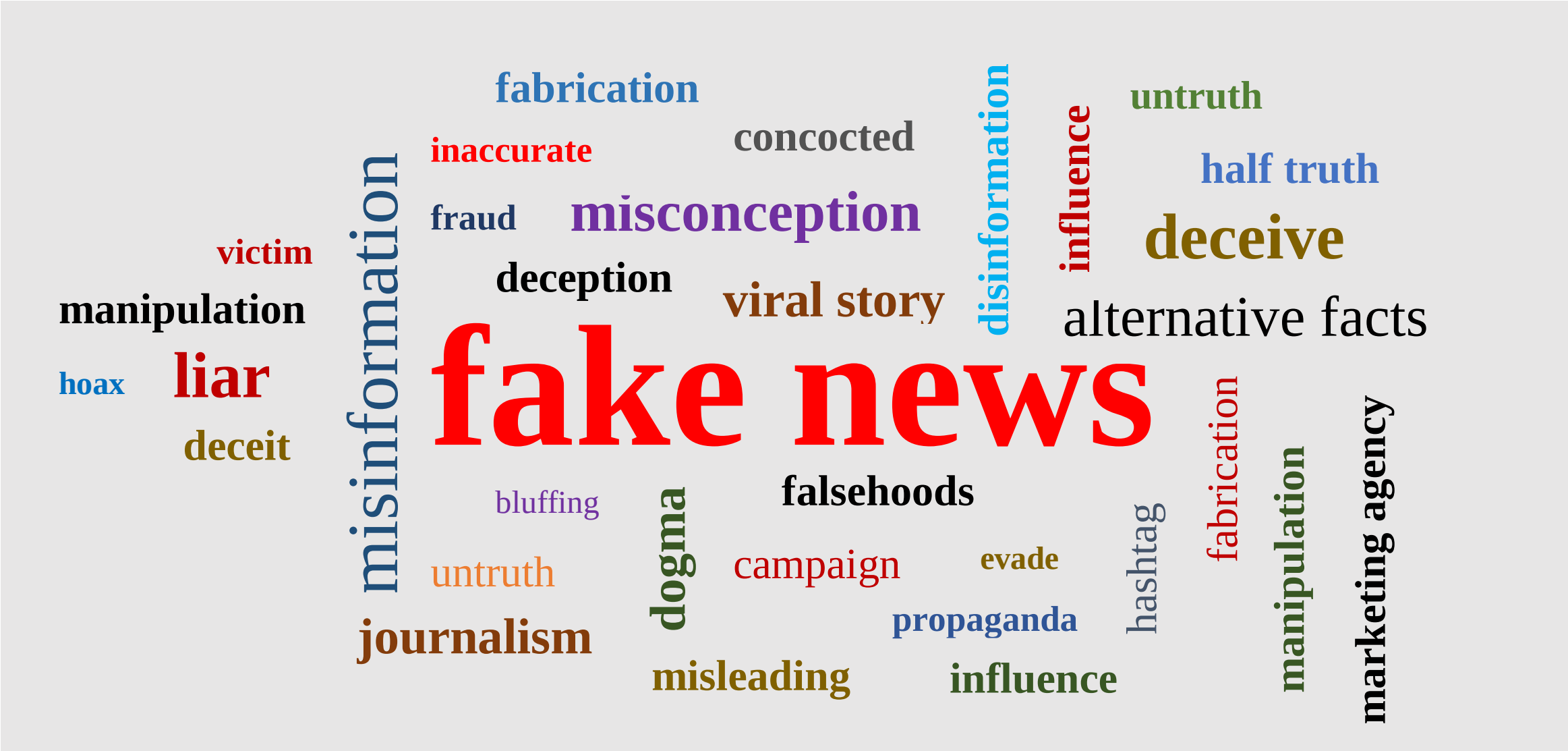

Figure 1: Word cloud representing different connotation of fake news

Graphs normally occur in various real-world problems employing cultural and social issues. Information that is accessible to us can be transformed in a structured way by means of connections between various objects to give promising insights regarding data [5]. One crucial issue in Artificial Intelligence problems is to draw meaningful relation between one part of data to other parts of data. Currently the best way to describe/represent these relations and connections between data is by employing graphs data structure that exist universally in modern technology solutions by providing us with a plan to mathematically describe complicated data as well as associated parameters.

Artificial Intelligence problem solving techniques can be broadly categorized into three classes: supervised, unsupervised and semi-supervised. Supervised technologies are well established to provide benchmark solutions in healthcare, data authentication, web services, security and many more fields. The realm of semi-supervised and unsupervised paradigm still needs to be judiciously explored. The major pitfall of supervised training methods are annotation inconsistencies, cost and manpower required due to massive data size. In order to create a good balance between labelled and unlabelled data we tried to use both at their best level of performance. Labelled data steers the training process towards right direction and unlabelled data is used to enhance model generalization as well as performance. Therefore, a good balance of both is highly required that we categorically attempted to incorporate in our suggested framework.

Kusner et al [6] established Word Mover's Distance in 2015 as a distinct case of Earth Mover's Distance [7] and also proved that WMD gives unprecedented performance when coupled with K-nearest neighbour in classification tasks. This measure is suitable for calculating the distance between two text documents based on alignment between the words that are semantically close but syntactically different. It basically measures the dissimilarity between two text documents as the minimum distance the word vectors of one document need to travel to reach the word vectors of another document. GCN and its variants are being well tested to provide outstanding performance in variety of real-life applications incorporating both textual and visual data such as automatic seizure detection using temporal GCN [8], spatio-temporal GCN for picking up and returning demand of public bike sharing by exploring information from multi-view data [9], Autism Spectrum Disorder and Alzheimer's Disease prediction with imaging and non-imaging information [10] , semi-supervised hyperspectral image (HIS) classification framework using spectral-spatial graph convolution network ($S^2$GCN) [11], multi-stream attention-enhanced adaptive graph convolutional neural network (MS-AAGCN) for skeleton-based action recognition [12] etc.

The major problem we face in classification techniques is less no. of available labels for data. Apart from this cost, manpower and labelling inconsistencies are major concerns for annotating huge volumes of news articles. Thus, we make use of semi-supervised learning method. Graphs are capable of representing structural and contextual dependencies between its nodes, each representing an object and subsequently can be utilized in cases where the number of labels is limited [13]. This encourages us to take benefits of Graph Convolution Network for classification. Given one node, for all the classes can virtually start training procedure. Such a model, can very quickly lead to the creation of a 2d vector space with word embeddings involved which could be visualized easily. As a result, it can be seen that with just 3 layers in the architecture along with one label per actual class and absolutely no features in place, a linear classification is made. This produces quite exceptional results with given constraints and conditions. The key contributions of proposed framework are as follows:

- The architecture presents a semi-supervised fake news detection technique based on GCN (Graph Convolutional Networks) trained with limited amount of labelled data.
- Elaboratively elucidate three building blocks of the framework: collecting word embeddings from the news articles in datasets utilising GloVe, constructing similarity graph using Word Mover's Distance (WMD) and finally applying Graph Convolution Network (GCN) for binary classification of news articles in semi-supervised paradigm.

- Graph Convolution Network can harness the best advantage of convolutions as well as data structuring capabilities of graphs to draw meaningful insights out of complicated data and associated parameters.
- The implemented technique is validated on three different datasets by varying the volume of labelled data.
- Experimental results are analysed for two different graph formulations constructed by taking k=3 and k=5 .
- Comparison with other contemporary techniques also reinforced the supremacy of the proposed framework.

The organisation of the work is done under five different sections and sub-sections. Section 1 highlights the problem of fake news circulation and its adverse effects on mankind. Significant works done in literature to eradicate the problem of veracity analysis and content authentication on web platforms is detailed in section 2. Section 3 deals with the technical specifications of the proposed framework and its associated components in different sub-sections. Implementation, experimentation, datasets, testing and contemporary comparison is being described in section 4. Finally, section 5 concludes the work with possible future directions to improve it.

## 2. Related Works

The subject of fake news becomes an emerging issue within modern social media soundings. Current veracity analysis methods can be broadly classified into a couple of sections [13], [14] : utilising news data for handling social information/context, extracting linguistic, syntactic, lexical features or visual characteristics [15], [16] from news articles to obtain particular features as well as interesting patterns that usually happen in fraudulent articles. Visual features are adopted to recognise fake pictures that are deliberately produced to exert realistic effects in fake news. Context-based approaches of veracity analysis includes characteristics of user-data on social sites and social systems [17]. User social site's profile is utilised to estimate that users' features and reliability. Characteristics derived from their social posts describe their social behaviour. Pathak et al. [18] presented elaborated discussions on characteristics of publicly available datasets for veracity analysis techniques and performance comparison with contemporary machine learning, deep learning, hybrid approaches of supervised and unsupervised domain. Contemporary survey of semi-supervised and unsupervised methods of web content credibility analysis is detailed by Saini et al [19]. We studied and categorized the available literature in the domain of veracity analysis into three different classes: supervised, semi-supervised and unsupervised methods. Apart from this also highlighted the unprecedented performance of different variants of GCN in a variety of real-life applications including healthcare, recommendation systems, traffic management, e-commerce, retail, marketing etc. that serves as a motivation to utilize it further for our designed fake news detection framework.

### 2.1 Supervised Methods

The supervised methods principally concentrate on selecting useful characteristics and utilise them to develop supervised training structures. A neural network-based Graph-aware Co-Attention Networks (GCAN) [20] framework considers realistic scenario of social network to predict whether a source tweet is fake or real at the same time producing the reasonable explanations as well as highlighting evidences on suspicious tweeters. The network requires the input in the form of short text tweets and the corresponding sequence of retweet users without text comments. GCAN significantly outperforms contemporary techniques by 16% in accuracy on average on Twitter 15 and Twitter 16 datasets. To model the sematic representations of fake news by jointly modelling the textual representations, knowledge concepts and visual information into a unified framework Wang et al. [21] proposed a novel Knowledge-driven Multimodal Graph convolutional Network (KMGCN). After knowledge distillation process a well-designed graph structure is used to effectively represent textual, knowledge and visual information to apply GCN succeeded by pooling layer, fully connect layer and finally softmax function for classification. The framework is experimented and validated on WEIBO and PHEME datasets. A weakly supervised reinforcement learning framework (WeFEND) proposed by Wang et al. [22] comprises of three components: annotator, reinforced selector and fake news detector. The annotator automatically assigns weak labels for news articles, the reinforced selector filters out high-quality samples using reinforcement learning and the fake news detector classifies fake articles based on content of news. Wu et al. [23] proposed a graph-based method of rumour detection in which a propagation graph is being constructed based on who replies to whom relationship on Twitter. Two models GLO-PGNN (global embedding with propagation graph neural network) and ENS-PGNN (ensemble learning with propagation graph neural network) adopting diverse classification approaches for rumour detection task are designed. Both the methods employ attention mechanism for dynamically adjusting the weights to improve performance. Meel and Vishwakarma [24] proposed multimodal fake news detection method based on ensemble of hierarchical attention network, image captioning and forensics analysis. Raj and Meel [25] experimented with Resnet50 , VGG16, VGG19,InceptionV3, DenseNet, Xception, AlexNet , MobileNet deep arcitectures on T1-CNN, EMERGENT, MICC-F220 multimodal datasets to compare and contrast the capacities of different ConvNet architectures for textual as well as visual forgery detection.

### 2.2 Semi-supervised methods

A semi-supervised naïve architecture Factual News Graph (FNG) [26] based on graphical social context representation is proposed that demonstrates significant improvements in fake news detection task and robustness in the presence of limited training data. The FANG representation learning framework is optimized by three concurrent losses unsupervised proximity loss, self-supervised stance loss and supervised fake news detection loss. Guacho et al. [27] recommended semi-supervised misinformation detection in graphs via embedding the input articles in multi-dimensional tensors which are further decomposed to capture spatial and contextual information by creating article-by-article-graph. A temporal ensembling based semi-supervised fake news detection framework using self-ensembling of the combined prediction of all previous epochs to serve as a proxy label for the current epoch is proposed by Meel and Vishwakarma [2] . Online opinion reviews have an economic impact on goods and services. Opportunistic firms often manipulate online reviews to make undue profits by unethically influencing people's choices. Rout et al. [28] proposed a semi-supervised method of classifying deceptive and fake opinion reviews and validated the framework on hotel reviews. A novel semi-supervised framework for multimodal fake news detection is designed by Mansouri et al. [29] using Linear Discrimination Analysis (LDA) and Convolutional Neural Network (CNN). Li et al [30] proposed a self-learning semi-supervised deep learning network that adds a confidence network layer to made it possible to automatically return and add correct results for improving the accuracy of the neural network.

### 2.3 Unsupervised methods

Annotated samples cannot characterize the authenticity of the news on recently developed event as they become obsolete very quickly due to the dynamic nature of news. To obtain high-quality up-to-the-minute labelled sample is the foremost challenge in supervised and semi-supervised learning methodologies. The motive behind development of unsupervised technologies is to completely eliminate the need of annotation and real time solution to the problem of information trustworthiness. A novel unsupervised method of detecting fake news circulating on WhatsApp is implemented by Gaglani et al. [31] leveraging Transfer Learning techniques. The framework utilizes semantic similarity between claims circulated on WhatsApp and associated articles scrapped from web to classify the claims as true or false. Hosseinimotlagh and Papalexakis [32] recommended an unsupervised ensemble method that consolidates results from different tensor decompositions into coherent, comprehensible and high precision groups of articles that belong to different categories of false news. Unsupervised microblog rumour detection framework is proposed [33] based on recurrent neural network and autoencoders by analysing the temporal dynamics and crowd wisdom. Yang et al. [34] recommended a generative approach of unsupervised fake news classification on social media by exploiting user's engagement, opinion and their credibility factors. A three phased graph-based approach utilizing biclique identification, graph-based feature vector learning and label spreading for fake news detection in the absence of annotated data is successfully designed by Gangireddy et al. [35].

### 2.4 Application areas of GCN

Detailed survey analysis of GCN in different real-life applications is being presented by Singh et al. [36]. Graph Convolution Network is being explored by Yao et. al [37] in conjunction with LSTM network to explore between objects in an image based on their spatial and semantic connections for generating image captions. Jeong et al. [38] proposed a deep learning based context aware citation recommendation framework encompasses a document encoder and content encoder using GCN and BERT .Other remarkable contributions of GCN for designing solutions to challenging problems can be highlighted as: semi-supervised hyperspectral image (HIS) classification framework using spectral-spatial graph convolution network ($S^2$GCN) [11], Infusing knowledge into the textual entailment tasks [39], multi-label image classification framework using GCN [40] , robot navigation in crowds using GCN with attention trained by human gaze data that accurately predicts human attention to different agents in the crowd [41] etc.

### 2.5 Motivation and Research Objectives

The inspirations for developing semi-supervised fake news detection framework and prime research targets can be regarded as follows:

- To design a semi-supervised Fake News Detection framework that can learn from the semantics of labelled data and intrinsic patterns of unlabelled data to optimize in terms of time, cost, labour and annotation inconsistencies.

- To address the highly circulated forgery format of a news story coupled with headline text along with partially labelled training samples.
- To take advantage of self-learning cognition of designed framework which facilitates the architecture to learn from up-to-the minute latest forgery formats fed into the process in the form of unlabelled articles.
- Kusner et al [6] established the fact that Word Mover's Distance gives unprecedented performance when coupled with K-nearest neighbour in classification tasks.
- Graph Convolution Network can harness the best advantage of convolutions as well as data structuring capabilities of graphs are capable of representing structural and contextual dependencies between its nodes (documents) to draw meaningful insights out of complicated data and associated parameters.
- The framework is aimed to avoid cumbersome feature engineering of supervised learning methods, crowdsourcing or human expert annotation inconsistencies, reduces cost and manpower requirements.

## 3. Proposed Model

The semi-supervised framework of Graph Convolutional Network combines best features of convolutions and data modelling capabilities of graphs. The proposed architecture incorporates a three-step approach to classify the articles as fake or real. First step is to transform the given textual data from a dataset in the form of vectors. These vectors represent the linguistic features of the text and can be utilized to represent an article in Euclidean Space. Global Vector (GloVe) embedding is used to transform the articles into vectors. Each article is interpreted as the mean of vectors of the words it contains. The result of this embedding is used to construct a similarity graph between articles in the dataset. Each node represents an article and most similar nodes are connected by an edge in the graph. To calculate the similarity, i.e. distance between two articles, Word Mover's Distance is used. It uses semantically meaningful relations between words to find the similarity between two articles. Before converting text into vectors basic pre-processing steps such as: dropping unused columns, removing null and missing data, removing stope words, tokenizing the articles, lemmatizing each word, converting labels name etc. are done for all the datasets. The subsequent procedure involves word embedding and similarity graph construction. This results in creation of graphs containing three types of nodes labelled as real, fake and unlabelled which can be inputs to the graph convolutional neural networks for finally predicting the output. The process is pictorially emphasized in figure 2 and also comprehensively detailed in algorithm1.

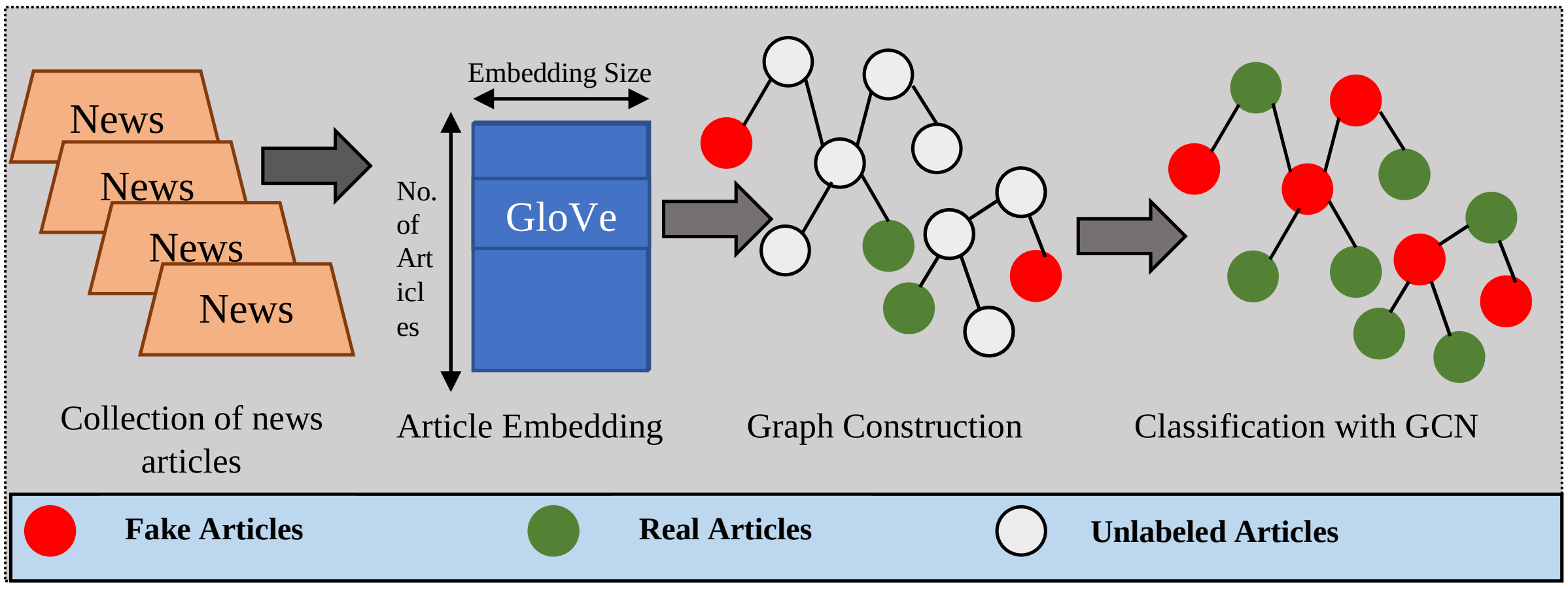

Figure 2: Proposed semi-supervised fake news classification framework

**Algorithm 1: Learning in Semi-supervised GCN Framework**

1: Perform initial pre-processing on input samples
2: Split the input samples into 80% training,10% validation,10% testing
3: Article embedding in Euclidean space using Glove with Embedding Dimension=300
4: Represent every article as the mean vector of the corresponding GloVe embedding
5: Similarity graph construction using WMD with K-nearest neighbour; K=3 and K=5
6: Train GCN of 4 layers up to 120 epochs with dropout=0.5 and learning rate=0.01
7: Evaluate the performance on test set

### 3.1 Article Embeddings in Euclidean Space

Article embedding in Euclidean Space means transforming text to multidimensional vectors by using word embedding. Word embedding is the process of constructing mathematical equivalents, i.e., representation in the mathematical form of each word from some corpus. It is implied that words with similar or same sense will have close representations. In Natural Language Processing, word embedding has proved to be one of the concepts that have huge applications, as it lays the foundation for solving many real-world tough problems. The process involves constructing vectors for each word in any number of dimensions such that it can be visualized as a vector in a vector space in those number of dimensions. This conversion of words to vectors uses neural network methodology to find the values of vectors. The vector can have any number of dimensions ranging from a single digit to a few hundred. As words with similar or same sense will have close representations, it can be implied that the meaning of words has been learnt by a model, which in turn can be useful for solving challenging problems. Our method will utilize the final word vectors created by the embedding of words for each article by representing it in an euclidean space. Further this will also be converted into low dimensions using a dimensionality reduction technique and contextual associations in news articles will be obtained by using a similarity graph representation.

To create a vector which maps to each news article, our proposed semi-supervised method uses a pre-trained GloVe model of 300 dimensional embeddings and for all the words appearing in that article we compute a mean vector which is finally mapped to that article. GloVe (Global Vectors) an unsupervised learning method was developed at Stanford university as an open-source project by Pennington et al [42] and was launched in 2014. Word Vectors in such an n-dimensional space, occurs in the manner as similar words appear close to each other while words with different meanings appear far from each other. One major advantage that GloVe offers is its dependency on global statistics instead of local statistics as used by Word2Vec or some other embedding techniques. GloVe is based on the basic principle of co-occurrence matrix. A simple example of a co - occurrence matrix of window size 1 is depicted in Figure 3, illustrates how words befall collectively that ultimately produces the relations between individual words. Co-occurrence matrix for words is calculated just by adding the mutuality of occurrence of words in an obtained text. In general use cases this co-occurrence matrix is decomposed by employing dimensional reduction techniques like PCA and SVD.

| | the | cat | sat | on | mat |
|---|---|---|---|---|---|
| the | 0 | 1 | 0 | 1 | 1 |
| cat | 1 | 0 | 1 | 0 | 0 |
| sat | 0 | 1 | 0 | 1 | 0 |
| on | 1 | 0 | 1 | 0 | 0 |
| mat | 1 | 0 | 0 | 0 | 0 |

Figure 3: An example co-occurrence matrix for the sentence “the cat sat on the mat”

### 3.2 Similarity Graph Construction

The classification step requires a graph input, which is done by constructing a similarity graph representing closeness in between different nodes, i.e., each news article. Such a graph can be represented as a k-nearest neighbour graph. In particular, for every news article, we look forward towards the k–nearest neighbours in the embedding space. The k neighbours can be determined by calculating the Word Mover’s distance between one article to all the other news articles.

A k-NN (k-nearest neighbour) graph is one in which node p and node q have an edge between them if node p is one of the top k nearest of all its neighbours or vice-versa. Every k-nearest-neighbours of a location in the n-dimensional space will be determined to utilise a "closeness" relationship wherever closeness is usually described within words of a distance measure as Word Mover's Distance. Thus, with given articles in a vector space, a k-NN graph of points can be created by calculating the remoteness between each pair of points and connecting each point with the k most proximal ones.

The Word Mover's Distance (WMD) proposed by Kusner et al. [6] in 2015 is a distance between documents that takes benefit of semantic relations among words that are apprehended by their embeddings. It is a special case of Earth Mover's Distance [7] and also provides extraordinary performance when coupled with K-nearest neighbour in classification tasks. It basically measures the dissimilarity between two text documents as the minimum distance the word vectors of one document need to travel to reach the word vectors of another document. This distance proved to be quite effective, obtaining state-of-art error rates for classification tasks.

WMD metric is a distance function between text documents that leads to unparalleled low k-nearest neighbour document classification error rate. Precisely representing the remoteness between two documents has far reaching utilities in document retrieval, multi-lingual document matching, text clustering etc. To calculate a distance between two text documents the basic unit is "travel cost" between two words. Let "A" and "B" be the representations of two text documents with |A| and |B| are the number of unique words in both the documents respectively. Each word $w_i$ in document A is allowed to be converted into any word in document B. $T_{ij} \geq 0$ denotes the amount of distance word $w_i$ in A has to travel to convert into word $w_j$ in B. T is a sparse transportation flow matrix. To transform A completely into B we have to confirm that the complete outgoing flow from i$^{th}$ word equals $A_i$ and incoming flow to j$^{th}$ word must match $B_j$ represented mathematically as equation 1.

$$\sum_j T_{ij}=A_i \text{ and } \sum_i T_{ij}=B_j \quad (1)$$

The WMD distance between two documents is the minimum weighted cumulative cost required to move all words from document A to document B represented as

$$\text{WMD (A, B)} = \sum_{ij} T_{ij}\, c(w_i, w_j) \quad (2)$$

Formally the WMD between two documents is defined as the value of the optimal solution of the following transportation problem which is a special case of Earth mover's distance.

$$\min_{T\geq 0} \sum_{i=1}^{|A|} \sum_{j=1}^{|B|} c(w_i, w_j)\, T_{ij} \quad (3)$$

$$\sum_{j=1}^{|B|} T_{ij} = A_i \qquad \forall\ i\in\{1, 2, 3\ldots\ldots|A|\} \tag{4}$$

$$\sum_{i=1}^{|A|} T_{ij} = B_j \qquad \forall\ j\in\{1,2, 3\ldots\ldots|B|\} \tag{5}$$

$$T_{ij} \geq 0 \qquad \text{for all i , j} \tag{6}$$

Word Mover's Distance (WMD) is derived upon current events in embeddings of the words which determine the semantically significant description of those words of local co-occurrences within those sentences among new articles. Salient features of Word Mover's Distance can be characterized as:

- It is easy to learn, practice and free from the effect of any hyperparameters.
- It can be easily described as the distance among two text contents that can be split and described as the distances which are sparse within different words.
- It commonly includes some information represented as the word2vec or Glove and heads to huge operations accuracy.

### 3.3 GCN for Graph Classification

Graph Convolutional Network is a resourceful development of convolutional neural networks which functions directly on graphs. The model scales linearly in the number of graph edges by using a competent layer-wise propagation rule that is based on first-order approximation of spectral convolutions on graphs. GCN model is capable of learning hidden layer representations that encodes both node feature and graph structure to an extent useful for semi-supervised classification.

Graph Convolutional Networks (GCN) are an ideological extension to Convolutional Neural Networks (CNN) where convolution process is applied on a graph instead of pixels which constitute the image. As CNN can capture information from the images and this information then can be used to classify images, a similar approach can be built over graphs as well. A filter analogous to a CNN filter can be employed in case of GCN to capture the similarity in graphs. Graph Convolutional Networks (GCN), like Convolutional Neural Networks (CNN) have bounded no. of hyper-parameters, which leads these techniques to occupy less memory and thus, multiple levels of information can be built to give a remarkable result when compared to a traditional learning algorithm. Figure 4 illustrates the basic design pipeline for a GCN model.

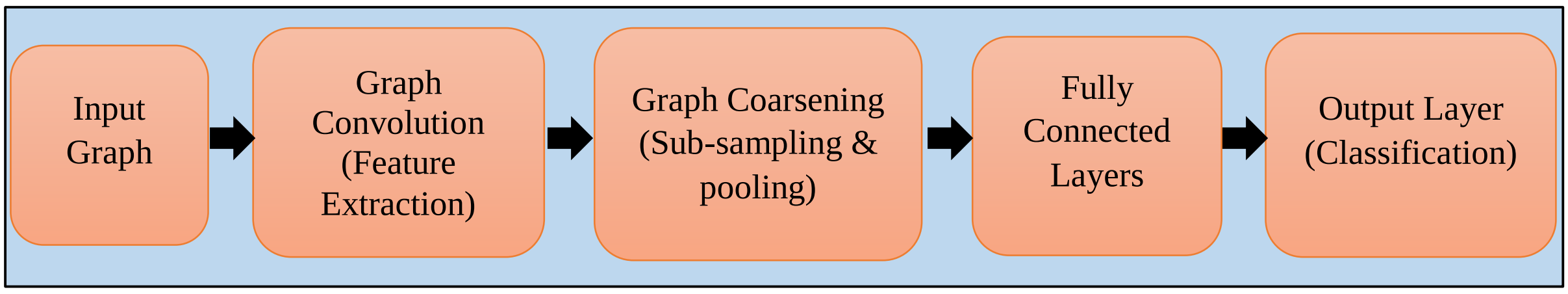

Figure 4: Design pipeline for GCN Model

The main purpose of graph classification is to foretell the labels of nodes within that particular graph. A node itself in the graph normally denotes a real-world object i.e. in our framework it represents news articles. The graph has been extensively employed to represent objects of real-world and the connection between them. Node/graph classification results prediction of labels of nodes in the graphs, for this purpose several researches take advantage of relationships among nodes to increase the classification efficiency.

GCN is one of the prominent variants of Graph Neural Network being used for a lot of real-life applications with non-Euclidean real world data having graph structure. The major difference between CNN and GNN is that GNN is a generalized variant of CNN built to operate on irregular or non-Euclidean structured data whereas the CNN can function only in case where the underlying data is regular (Euclidean). GCN is developed by Thomas Kipf and Max Welling [43] in 2016. Convolution in GCN is almost the same operation as the convolution in CNN. It implies multiplying the input neurons with a set of weights known as kernels or filters acts as sliding window and enables the network to learn features from neighbouring cells. Different layers may contain filters of different weights but within the same layer same filter will be used known as weight sharing. The particular variant of GCN that we use in our framework for semi-supervised classification is Spectral Graph Convolutional Network proposed by Thomas Kipf and Max Welling.

Neural networks apply non-linear activation functions to represent the non-linear features in latent dimensions. Forward pass equation in neural network is represented as in equation (7) where σ represents activation function, $H^{(l+1)}$ and $H^{(l)}$ represents feature representation at l$^{th}$ and (l+1)$^{th}$ layer, $W^{(l)}$ weight and $b^{(l)}$ bias at layer l.

$$H^{(l+1)} = \sigma(W^{(l)} H^{(l)} + b^{(l)}) \quad (7)$$

Forward pass equation in Graph Convolutional Network can be represented as:

$$H^{(l+1)} = \sigma(W^{(l)} H^{(l)} A) \quad (8)$$

A is adjacency matrix representing the connections between the nodes in graph structured real world data. Adjacency matrix A enables the model to learn the feature representations based on nodes connectivity. Bias b is omitted to make the model simpler. Equation (8) is the first-order approximation of spectral graph convolution propagating the information along the neighbouring nodes within the graph.

The aim of Graph Convolutional Network framework is to learn a function of features on a graph which takes as input a feature matrix X of dimension NxD where N is the number of nodes and D is number of input features. A is the adjacency matrix for representing the overall graph structure. A and X are the input to GCN architecture producing the node level output Z (an NxF feature matrix, where F is the number of output features per node). A nonlinear function representing every neural network layer in GCN can be written as in equation (9) and equation (10) highlights layer wise propagation across the network.

$$H^{(l+1)} = f(\, H^{(l)}, A) \tag{9}$$

$$f\big(\, H^{(l)}, A\big) = \sigma(A\, H^{(l)}\, W^{(l)}) \tag{10}$$

With $H^{(0)}$=X is initial input, $H^{(L)}$=Z is node level output at last layer (or z for graph level output) , L is the number of layers , $W^{(l)}$ is a weight matrix for the $l^{th}$ neural network layer and σ is a non-linear activation function. Instead of using matrix A for semi-supervised classification in practical scenario Kipf and Welling [43] proposed to use symmetric normalization $D^{-1/2} AD^{-1/2}$, so equation (9) can be rewritten as:

$$f\big(\, H^{(l)}, A\big) = \sigma(\, \widehat{D}^{(-1/2)} \hat{A}\, \widehat{D}^{-1/2}\, H^{(l)}\, W^{(l)}) \tag{11}$$

With $\hat{A}$ = A+I, I is identity matrix and $\widehat{D}$ is the diagonal node degree matrix of A. In the process of semi-supervised classification with GCN the network initially starts training on the labelled nodes, subsequently propagating the information to unlabelled nodes by updating weight matrices that are shared across all nodes. This process can be summarized in the following steps:

- Accomplish forward propagation through GCN.
- Put on the sigmoid function row-wise on the last layer in the GCN.
- Calculate the cross-entropy loss on known node labels.
- Backpropagate the loss and update the weight matrices ‘W’ in each layer.

# 4. Experiments

In this sub-section, we extensively detail the experimentation settings, datasets, parameter selection and insights gained from the wide range of experiments performed. The implementation is done on Google Colab which offers up to 13.53 free RAM and 12 GB NVIDIA Tesla K80 GPU. The proposed framework is built and implemented in Python 3 on top of the Keras deep learning framework. To make our implementation better and effective the experiments are repeated for different volumes of labelled and unlabeled data samples. The performance scores and comparison of the proposed framework are listed in terms of F1-measure, accuracy, recall and precision evaluation metrics. Analysis of the results is being done in numerical as well as in graphical representation using the shapes of accuracy-loss curves with epochs and area under the curve plots individually for each dataset.

After initial pre-processing each dataset is split into 8:1:1 ratio for training, validation and testing respectively. Articles are embedded in vector space using 300-dimensional GloVe word embedding. Word Mover's Distance is employed for constructing similarity graph utilizing K-nearest neighbours for two different values of K i.e. K=3 and K=5. Graph convolutional network involves 4 layers convolutions for feature extraction with intermediate leaky ReLU activation and dropout layers. GCN of 4 layers is trained up to 120 epochs with dropout 0.5 , learning rate 0.01 , 16 hidden units and weight decay of 5^e-4 . After fully connected layers in output layer activation function is softmax for binary classification of news instances into real or fake.

## 4.1 Datasets

Three different fully labelled datasets Fake News Data, Real or Fake and Fake News Detection introduced on Kaggle platform are used for experimentation and validation of our work. The headline, body and label part of each one of the datasets are being utilized for model training and testing purpose. The details of the datasets are listed in following Table 1:

Table 1: Dataset Details

| **Datasets** | **Details** | **Attributes Used** | **Total entries** | **Fake news count** | **Real news count** |
|---|---|---|---|---|---|
| Fake News Data [44] | Hosted on Kaggle, contains id, Title, Author, Text and Label | Headline/Title, Body/Text, Label | 20700 | 10360 | 10340 |
| Real or Fake [45] | Hosted on Kaggle platform, contains four fields Id, Headline, Body , Label | Headline/Title, Body/Text, Label | 6000 | 3000 | 3000 |
| Fake News Detection [46] | Hosted on Kaggle, contains site URL, Headline, Body and Label | Headline/Title, Body/Text, Label | 3988 | 2121 | 1867 |

Fake News Data [44] was presented on Kaggle website for Kaggle competition two years ago, now available publicly with annotations for research and learning purpose. It comprises of 20, 800 instances with five attributes: Id, Title, Author, Text and Label. After initial pre-processing, we have 20, 700 entries in the dataset with 10360 Fake news and 10340 real news entries. We utilized Title, Text and label features for our model training, testing and comparison purpose.

Real or Fake [45] dataset was introduced on Kaggle platform three years ago, now is being extensively used for research purpose. It contains four attributes Id, Headline, Body and Label out of these four we have used headline, body and label part in our research. The dataset initially has 6335 entries reduced to 6000 after preliminary data pre-processing. Segregation of dataset is 3000 real and 3000 fake news instances.

Fake News Detection [46] dataset is compiled on the Kaggle website by Jruvika. It contains four attributes site URL, Headline, Body and Label (Real/Fake). The dataset initially contains 4009 news instances. After initial data cleaning such as removing the entries with misplaced labels, missing headline and body we have 3988 rows with 2121 Fake and 1867 Real news samples.

### 4.2 Result Analysis

Performance of the proposed framework on three different datasets is being evaluated and compared for accuracy, precision, recall and F1 score. The experiments for each dataset have been repeated for two different values of K (K=3, K=5) and varied proportions of labelled training data ranging from 20% to 50%. Table 2 ,3 and 4 highlights the results obtained out of experiments for Fake News Data, Real or Fake and Fake News Detection datasets respectively. Figure 5 ,6 and 7 enlightens the Accuracy-Epoch curve, Loss-Epoch curve and ROC curves for each one of the datasets.

Table 2: Result analysis on Fake News Data dataset

| **% Labelled data** | **K=3** | | | | **K=5** | | | |
|---|---|---|---|---|---|---|---|---|
| | Accuracy | Precision | Recall | F1-score | Accuracy | Precision | Recall | F1-score |
| 20% | 79.29 | 82.52 | 74.65 | 78.39 | 66.03 | 61.00 | 79.27 | 68.94 |
| 30% | 83.26 | 86.07 | 77.32 | 81.46 | 74.66 | 85.32 | 72.44 | 78.35 |
| 40% | 87.99 | 75.99 | 88.60 | 81.81 | 82.36 | 69.02 | 87.45 | 77.15 |
| 50% | **91.18** | **93.05** | **94.27** | **93.65** | 88.67 | 81.71 | 89.26 | 85.32 |

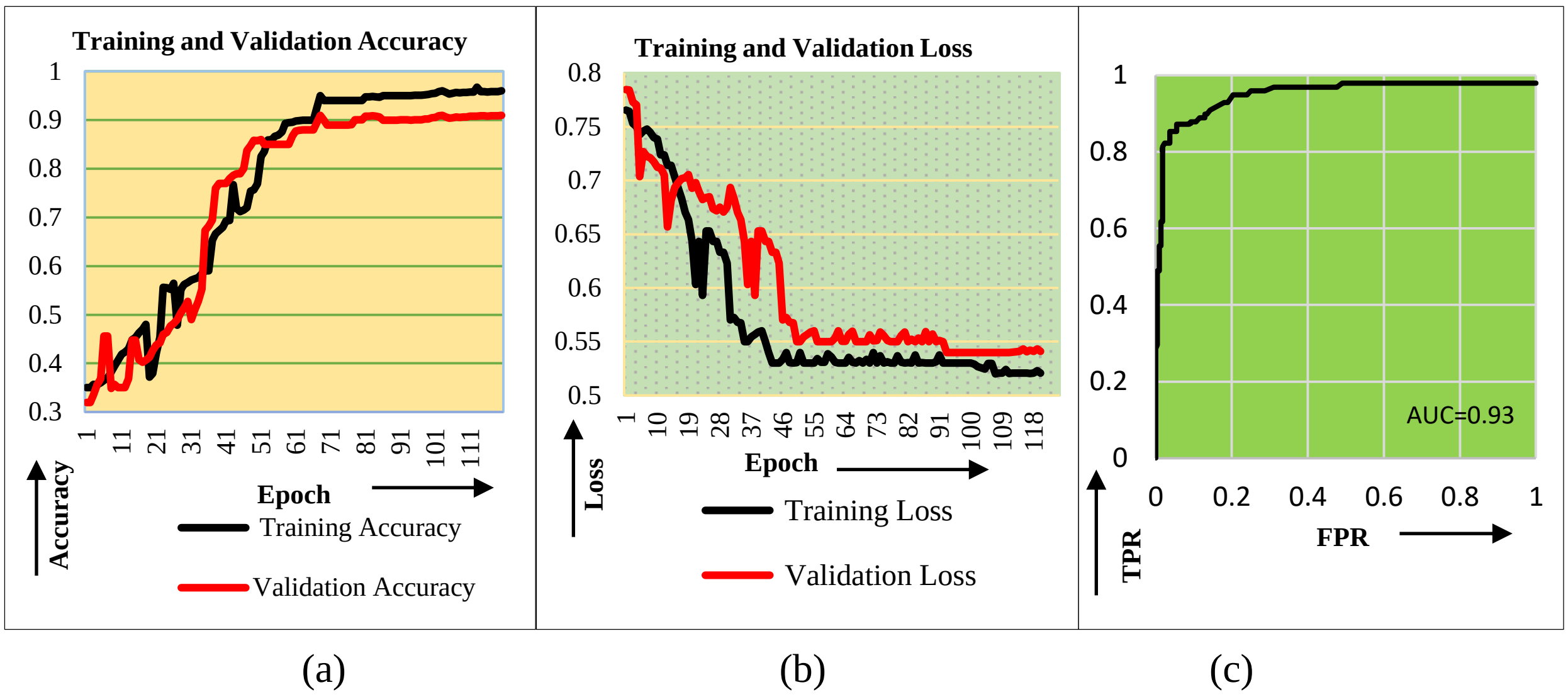

Figure 5: (a) Accuracy-Epoch curve (b) Loss-Epoch curve (c) ROC curve for Fake News Detection dataset

Table 3: Result analysis on Real or Fake dataset

| **% Labelled data** | **K=3** | | | | **K=5** | | | |
|---|---|---|---|---|---|---|---|---|
| | Accuracy | Precision | Recall | F1-score | Accuracy | Precision | Recall | F1-score |
| 20% | 79.55 | 83.75 | 79.99 | 81.83 | 80.00 | 83.69 | 70.08 | 76.28 |
| 30% | 86.32 | 75.00 | 86.76 | 80.45 | 83.33 | 75.00 | 79.27 | 77.08 |
| 40% | 91.02 | 89.00 | 90.23 | 89.61 | 87.19 | 76.68 | 93.55 | 84.28 |
| 50% | **95.27** | **89.47** | **95.99** | **92.61** | 92.34 | 84.29 | 92.52 | 88.21 |

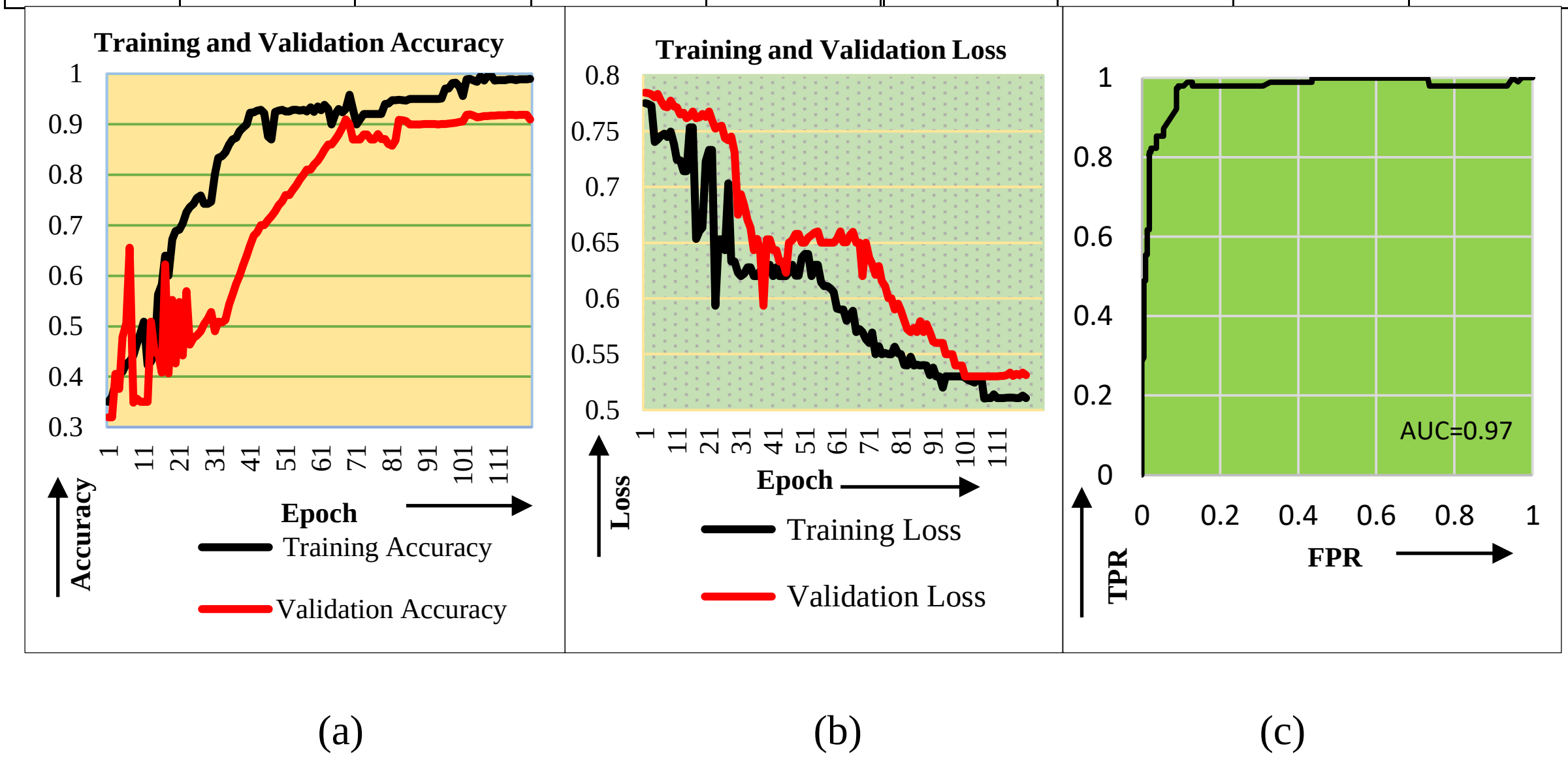

Figure 6: (a) Accuracy-Epoch curve (b) Loss-Epoch curve (c) ROC curve for Real or Fake dataset

Table 4: Result analysis on Fake News Detection dataset

| % Labelled data | K=3 | | | | K=5 | | | |
|---|---|---|---|---|---|---|---|---|
| | Accuracy | Precision | Recall | F1-score | Accuracy | Precision | Recall | F1-score |
| 20% | 75.03 | 76.06 | 71.27 | 73.58 | 77.32 | 81.33 | 76.65 | 78.92 |
| 30% | 81.26 | 75.60 | 86.59 | 80.72 | 86.23 | 87.88 | 78.19 | 82.75 |
| 40% | 85.04 | 88.27 | 87.77 | 88.02 | 88.69 | 86.01 | 86.62 | 86.31 |
| 50% | **92.03** | **92.07** | **94.87** | **93.45** | 90.37 | 83.33 | 90.17 | 86.61 |

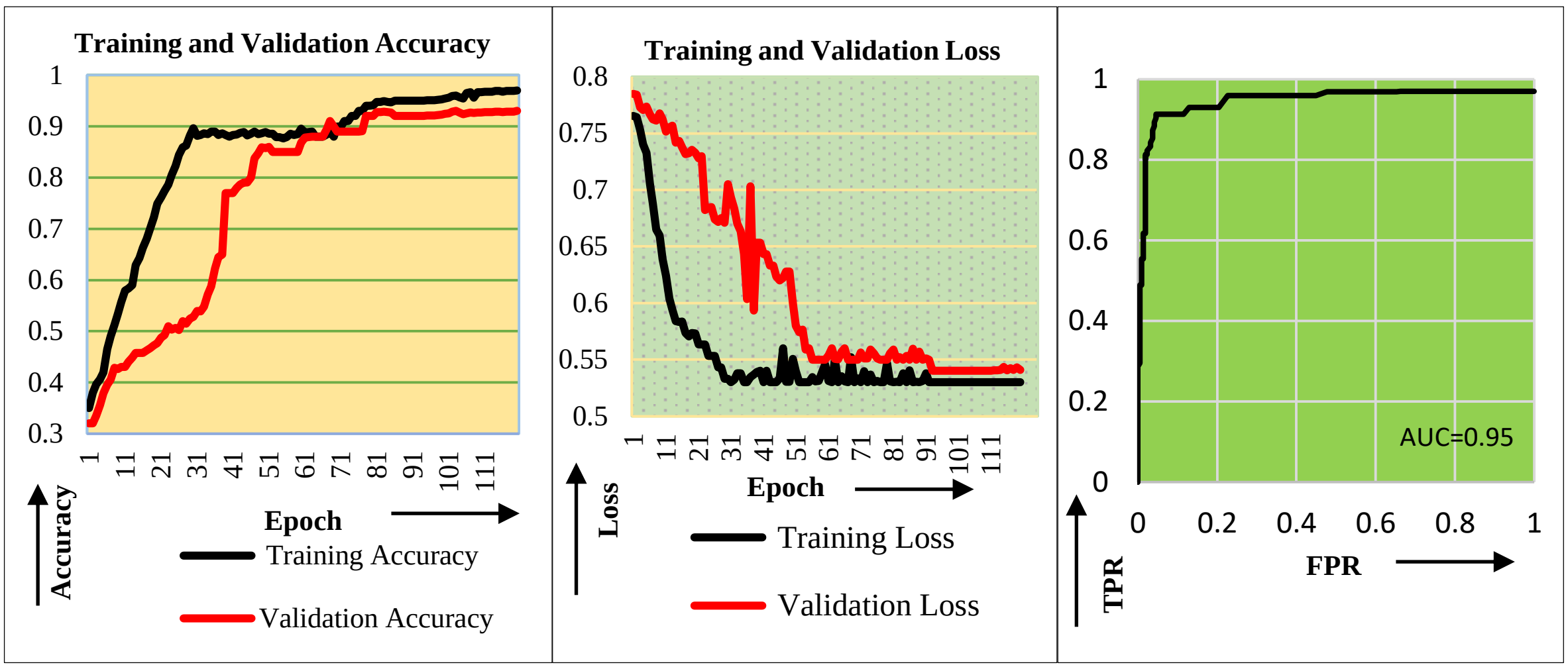

(a) (b) (c)

Figure 7: (a) Accuracy-Epoch curve (b) Loss-Epoch curve (c) ROC curve for Fake News Detection dataset

It is evident from the results analysed from Table 2, 3 and 4 that K=3 provides better accuracies than K=5 and as we increase the amount of labelled data in training process precision of detecting fake news also increases. The highest accuracy obtained from Fake News Data dataset is 91.18%, Real or Fake dataset is 95.27 % and Fake News Detection dataset is 92.03% respectively. Graphical representation of the experimental process in Figures 5, 6 and 7 advocates the overall effectiveness of the framework in terms of quality and quantity of performance.

### 4.3 State-of-the-art Comparison

To compare the efficacy of our designed architecture with contemporary methods, we calculated the performance of five different baseline methods in terms of accuracy, precision, recall and F1 score with the dataset split as 8:1:1 for training, validation and testing. State-of-the-art juxtaposition on each of the Fake News Data, Real or Fake and Fake News detection datasets are outlined in Table 5, Table 6 and Table 7, correspondingly. The approaches used as baselines for state-of-the-art comparison are as follows:

- Bali et al. [47] proposed Support Vector Classifier (SVC), Random Forest(RF), Naïve Bayes (NB), Multi-Layer Perceptron (MLP), K-Nearest Neighbour(KNN), AdaBoost (AB) and Gradient Boosting (XGB) methods by extracting sentiment polarity, 50-dimensional GloVe word embedding , n-gram count, and cosine similarity features between title and text part of news articles.

- Agarwalla et al. [48] applied Punkt statement tokenizer from NLTK library with Support Vector Machine (SVM), Logistic Regression and Naïve Bayes with Lidstone smoothing for classification.
- Karimi and Tang [49] developed a Hierarchical Discourse Level Structure using Bi-LSTM, which extracts structure-related properties of articles by building dependency trees.
- Vishwakarma et al. [50] suggested a framework of keyword extraction with web scrapping using a rule-based classifier. The classifier uses a reality parameter (Rp) which is considered by checking the credibility of the top 15 Google search results.

Table 5: Comparative performance analysis on Fake News Data dataset

| **Method** | **Accuracy (%)** | **Precision (%)** | **Recall (%)** | **F1-score (%)** |
|---|---|---|---|---|
| Bali et al. [47] | 91.05 | 93.00 | 94.00 | 93.49 |
| Agarwalla et al. [48] | 86.32 | 85.89 | 81.26 | 83.51 |
| Karimi and Tang [49] | 85.90 | 88.80 | 80.70 | 84.56 |
| Vishwakarma et al. [50] | 85.00 | 95.04 | 80.00 | 86.87 |
| **Proposed Method** | **91.18** | **93.05** | **94.27** | **93.65** |

Table 6: Comparative performance analysis on Real or Fake dataset

| **Method** | **Accuracy (%)** | **Precision (%)** | **Recall (%)** | **F1-score (%)** |
|---|---|---|---|---|
| Bali et al. [47] | 93.03 | 92.00 | 92.44 | 92.22 |
| Agarwalla et al. [48] | 90.00 | 89.90 | 90.00 | 89.94 |
| Karimi and Tang [49] | 89.06 | 82.90 | 88.70 | 85.70 |
| Vishwakarma et al. [50] | 91.01 | 90.50 | 95.70 | 93.03 |
| **Proposed Method** | **95.27** | **89.47** | **95.99** | **92.61** |

Table 7: Comparative performance analysis on Fake News Detection dataset

| **Method** | **Accuracy (%)** | **Precision (%)** | **Recall (%)** | **F1-score (%)** |
|---|---|---|---|---|
| Bali et al. [47] | 86.20 | 92.00 | 92.00 | 92.00 |
| Agarwalla et al. [48] | 83.87 | 81.22 | 82.67 | 81.93 |
| Karimi and Tang [49] | 83.09 | 80.22 | 82.01 | 81.10 |
| Vishwakarma et al. [50] | 88.30 | 85.20 | 88.40 | 86.77 |
| **Proposed Method** | **92.03** | **92.07** | **94.87** | **93.45** |

The above discussion concludes that our proposed model for veracity analysis of web information is reasonably promising. The precision over all the three datasets Fake News Data, Real or Fake and Fake news detection is a decent development over the parallel methods. Graph Convolutional Network and Word Mover's Distance for calculating distance are the prominent technologies that have facilitated refining the preciseness of our results. Finally, experimenting with different values of K and varied proportions of labelled data helps us achieve 95.27% highest fake news detection accuracy.

## 5. Conclusion and Future Works

Fake news is an increasing concern for the modern society. One of the most difficult part in providing a solution to this is labelling massive volumes of data to train supervised artificial intelligence models. To counter this challenge, we tried to design a semi-supervised text fake news detection framework based on Graph Convolutional Networks. Embedding the text articles in Euclidean space, similarity graph constructing using Word Mover's Distance and Graph Classification are the three landmark components of the designed architecture. Extensive experimental analysis has been done by repeating the training and testing of the model for different proportions of labelled and unlabelled data for each dataset. The promising results compared in terms of accuracy, precision, recall, F1 score and ROC Curves very well advocates the worthiness of the method for veracity analysis. The proposed framework exhibits encouraging results by outdoing numerous state-of-the-art methods on multiple standard datasets.

In spite of all the efforts done by researchers, governments and corporates to put an end to this malice of fake news, a lot more is still to be done. Temporal component is extremely important in current real-life decision-making process. Correct information received too late is also irrelevant, so real time information authentication systems are highly in demand. Platforms such as Facebook, WhatsApp, LinkedIn, Gmail etc. are implementing solutions for user privacy as well as data security and trustworthiness. The research could be further extended to incorporate multimedia data as visual data constitute an integral part of digital online information. Advanced deep learning technologies of transformers and transfer learning can also be amalgamated with graph data structures to make the process more efficient. Stand-alone modules that can be added as browser and application plugins could also be the subsequent research tracks. The classification performance of fake news detection systems can also be improved by using attention mechanism instead of complex convolutions as well as development of unsupervised frameworks on large scale unlabelled data further enhances the worthiness of the procedure.